\documentclass[twocolumn,prl]{revtex4}
\usepackage{graphicx}
\usepackage{dcolumn}
\usepackage{bm}
\usepackage{amsmath}
\usepackage{times}
\usepackage{epstopdf}
\usepackage{color}
\usepackage[breaklinks=true,colorlinks,citecolor=blue,linkcolor=blue,urlcolor=blue]{hyperref}

\makeatletter

\newcommand{\Rmnum}[1]{\expandafter\@slowromancap\romannumeral #1@}
\makeatother

\begin{document}

\title{Evidence of density waves in single crystalline nanowires of Pyrochlore Iridates}
\author{Abhishek Juyal}
\email{abijuyal@iitk.ac.in}
\author{Amit Agarwal}
\email{amitag@iitk.ac.in}
\affiliation{Department of Physics, Indian Institute of Technology Kanpur, Kanpur 208016, India}
\author{Soumik Mukhopadhyay}
\email{soumikm@iitk.ac.in}
\affiliation{Department of Physics, Indian Institute of Technology Kanpur, Kanpur 208016, India}

\begin{abstract}
We present experimental evidence of emergent density wave instability in single crystalline low dimensional wires of Yittrium based Pyrochlore Iridates. We demonstrate electric field induced nonlinear hysteretic switching of the density wave at low temperature, followed by smooth nonlinear conduction at higher temperature ($T > 40$ K) in Y$_{2-x}$Bi$_x$Ir$_2$O$_7$ with $x= 0$ and $0.3$. AC transport measurements reveal the presence of four different collective relaxation processes which dominate at different temperature scales. There is a strong coupling of the normal charge carriers with the density wave condensate, which is reflected in the linear scaling of the dc conductivity with the collective relaxation rate across a wide range of frequency and temperature regime. The evidence of density wave in low dimensional single crystals of Pyrochlore Iridate could be a precursor to the possible experimental confirmation of the Weyl semimetallic ground state with broken chiral symmetry.
\end{abstract}


\maketitle

\section{Introduction}
The physics of 5d heavy transition metal oxides (TMO) includes, \textit{inter alia}, a strong spin orbit interaction (0.1-1 eV) and its interplay with Coulomb correlations.
Among the 5d TMOs, rare earth Pyrochlore Iridates R$_2$Ir$_2$O$_7$ (R = Yttrium, or lanthanide elements) provide a promising experimental test bed to explore the realization of a
variety of novel quantum phases~\cite{Wan, Kim, Pesin} such as topological Mott insulators~\cite{Pesin, Yang, Kargarian}, chiral spin liquids~\cite{Machida}, Weyl semimetals (WSM)~\cite{Wan,
William}, and axion insulators~\cite{Wan, Go}, due to the competition between spin-orbit interaction, electron-electron interaction, and the kinetic hopping parameter.
Interestingly, chiral symmetry breaking induced by interaction effects in Weyl semimetals can also lead to charge density waves \cite{Axion-CDW}.  
However, from an experimental viewpoint, there is very little evidence to either support or disprove such claims.

Earlier experimental studies on Pyrochlore Iridates have shown the existence of a continuous Mott insulator-metal transition with increasing lattice spacing induced either by change in the ionic radius of R \cite{Matusihara} or by substitutional doping \cite{Aito}, along with evidence of strong correlation without any hard gap at the Fermi level~\cite{Singh}. Similar to the ruthenium Pyrochlore systems, this metal-insulator transition in Y$_{2-x}$Bi$_x$Ir$_2$O$_7$ is also accompanied by a magnetic phase transition with a small but finite value of spontaneous magnetic moment below the transition temperature \cite{Aito,Taira}. The magnetic ground state in some Pyrochlore Iridates is  believed to be spin-glass like \cite{Aito,Taira} as in ruthenium Pyrochlores, which develops into a long ranged  (probably noncollinear) magnetic ordering pattern
at low temperatures \cite{Wan, Disseler}, or into a quantum spin liquid as in Pr$_2$Ir$_2$O$_7$ \cite{Pr_SL}, although the precise nature of the ordering is not yet fully established~\cite{Wan, Shapiro}. The rich variety of quantum phases are most likely to be realized in mesoscopic
single crystalline systems. However, till date, experimental reports on Pyrochlore Iridates are mostly confined to bulk single crystalline or polycrystalline samples~\cite{Pr_SL, Dwivedi} and a few reports on epitaxial thin film growth of iridates~\cite{Fujita, Gallagher}. For thin films, extraneous factors such as lack of crystallinity, strain, defects, etc. could come in the way of a clear interpretation of the experimental data~\cite{Gallagher}.

\begin{figure}
\includegraphics[width=1.0\linewidth]{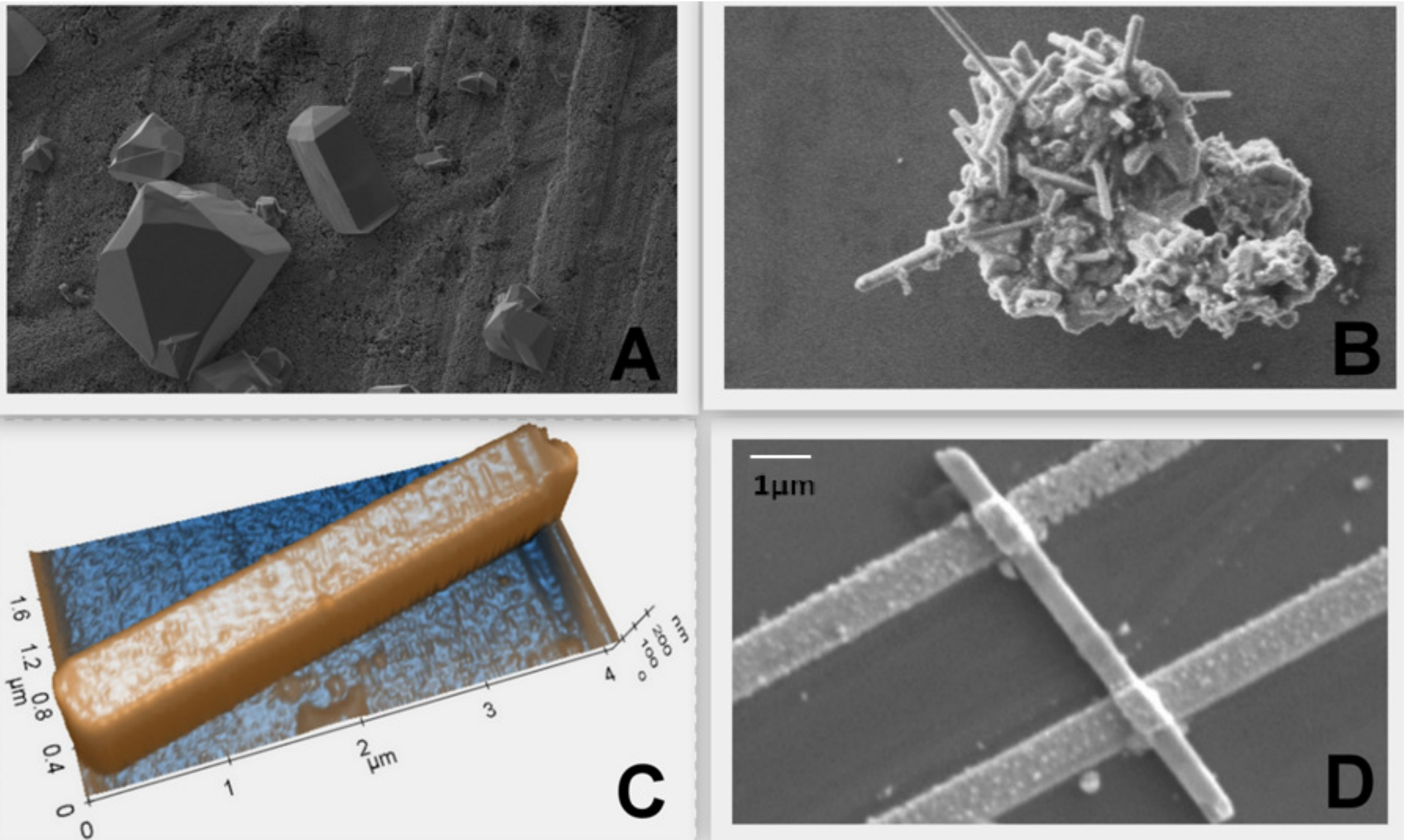}
\caption{A) SEM picture of faceted single crystals of YBIO, loosely attached to the poly-crystalline sample surface in contact with air.  B) Cylindrical growth of YIO wires from poly-crystalline YIO drop casted on a substrate after annealing at 800~$^0$C. C) AFM image of a single crystal wire of YBIO isolated from the bulk sample and transferred on a substrate.  D) Representative SEM image of a device with contacts made on the YBIO wire using e-beam lithography.}\label{fig1}
\end{figure}

\begin{figure*}[ht]
\includegraphics[width=0.7\linewidth]{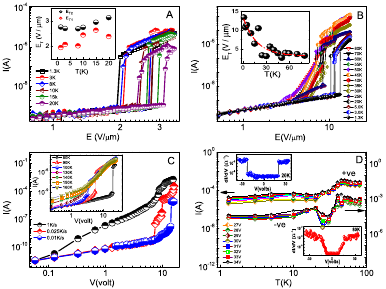}
\caption{A) I-V characteristics of a YBIO single crystal wire (200 $\times$ 400 nm rectangular cross section) showing hysteretic switching transition at low temperature. The inset shows the depinning threshold field for
increasing dc bias (labeled as $E_{\rm T1}$) which is always greater than that for the decreasing bias (labeled as $E_{\rm T2}$).
B) I-V characteristics of YIO single crystal wire (200 nm diameter)  displaying sliding transition of CDW with switching at low $T$ and non-switching behavior at relatively higher $T$. Note that the switching behavior at low $T$ ($T\le 40$ K) is hysteretic (the decreasing field cycle is not shown in the figure for clarity). Inset: The de-pinning threshold for the increasing dc bias $E_{\rm T} = E_{\rm T1}$ for low temperatures, far away from the CDW transition temperature, shows an exponential temperature dependence: $E_{\rm T} \propto \exp[-T/T_0]$, with $T_0 = 17$ K. C) A representative plot of the cooling rate dependence of de-pinning threshold for a $2\mu$m diameter YIO wire device. Inset: The dc I-V characteristics at different temperatures. D) Representative data for the temperature dependence of the current carried by the sliding CDW in YIO at different voltage levels above the de-pinning threshold for positive and negative biases are shown. The switching and non-switching regime is separated by a resistive anomaly (at $T \approx 40 $ K).
\label{fig2}}
\end{figure*}

In this article, we present dc and ac electrical transport study of devices fabricated from {\it single crystalline} nanowires of YBIO (Y$_{2-x}$Bi$_x$Ir$_2$O$_7$ with $x=0.3$) and YIO (Y$_2$Ir$_2$O$_7$). Our study clearly
establishes that the electronic ground state of these low dimensional systems is a charge density wave (CDW). CDWs are ordered states with broken translational symmetry, having periodic modulations of the electronic charge density accompanied by the associated lattice distortion in systems of electronically reduced dimensions~\cite{Gruner_rmp}. The density modulation is
triggered by a energy lowering mechanism; a Fermi surface nesting instability (also called Peierls instability) connecting the electron like and the corresponding hole like part, thus opening up a gap at the Fermi surface~\cite{Peierls,EPC}. When driven by an external dc electric field, the CDWs are `pinned' below certain threshold field ($E_{\rm T}$) due to presence of impurities or increased surface to volume ratio. Below $E_{\rm T}$, the transport is ohmic with contribution arising only from `normal' quasi-particles thermally excited above the CDW gap while above $E_{\rm T}$, the CDW slides relative to the host lattice with the collective motion leading to strong non-linear transport, with or without hysteretic switching at low temperatures ~\cite{Gruner_rmp}.

\section{Experimental results and discussion}
For this study, we extracted small faceted crystals from the bulk poly-crystalline sample surface of YBIO (Fig.~\ref{fig1}A) by exfoliation. The wire shaped crystals were transferred on a Silicon oxide substrate with pre-patterned alignment marks. Atomic force microscopy shows smooth sub-nanometer crystal surface roughness and rectangular cross-section (Fig.~\ref{fig1}C). After locating a suitable crystal, electron-beam lithography was used to form Au/Cr contact pads for electrical measurements (Fig.~\ref{fig1}D). For YIO, we observed no such crystal formation on the surface of the pelletized bulk sample. However, sparse quantities of YIO crystals were found in the polycrystalline sample sintered in powder form (Fig.~\ref{fig1}B). Typical length of these wires varies between 1-50 $\mu$m and the diameter is in the range 200nm-2$\mu$m. A solution containing suspended YIO crystals were prepared and drop casted on flat silicon oxide substrates, followed by metallization using e-beam lithography. The chemical composition of the crystals of YBIO and YIO was analyzed using energy dispersive X-ray spectroscopy (EDX) and was found to be similar to the respective bulk poly-crystalline samples. The details of sample preparation and characterization are given in Ref.~\cite{SM}.

\subsection{DC response and depinning of density waves}
The dc I-V characteristics data of two representative devices based on YBIO and YIO, respectively, is shown in Fig.~\ref{fig2}. Depending on the temperature, the dc I-V characteristics for both YBIO and YIO can be divided into certain distinct regimes. At very low temperature (typically $T < 40$ K), we find 2 regimes: 1) Ohmic behavior up to certain field value $E_{\rm T}$; 2) Beyond $E_{\rm T}$, there is a sharp hysteretic switching into a highly conducting non-linear state. On the other hand, at relatively higher temperature, the hysteresis disappears leading to 3 distinct regions in the I-V characteristics: 1) a Ohmic zone below $E_{\rm T}$; 2) appearance of non-linearity above $E_{\rm T}$; 3) non-hysteretic switching into another highly conducting nonlinear state above $E_{\rm G}$ ($E_{\rm G} > E_{\rm T}$). At still higher temperature, there are 2 regimes: Ohmic state followed by a highly nonlinear enhancement of conductance above $E_{\rm T}$ without any switching. We also find that the current carried by CDW is weakly temperature dependent with a resistive anomaly separating the switching regime and the non-switching regime near $T \sim 40$K -- see  Fig.~\ref{fig2}D. The differential conductivity (dI/dV) curve shown in the inset of Fig.~\ref{fig2}D, simply highlight the switching behavior of the CDW at low T, and non-switching behavior at higher T. However the CDW persists even beyond the switching regime even for higher T, as seen in the non-linear conductivity in Fig.~\ref{fig2}A-B.

From a theoretical perspective, the dynamics of a charge (or spin) density waves is effectively described by the standard Fisher-Lee-Rice (FLR) model ~\cite{Fukuyama}, which treats the CDW as a charged elastic object with an infinite internal degrees of freedom, deformable around the impurities. The competition between the random impurity potentials trying to pin down the phase of the complex order parameter of the CDW ($\Delta (r,t) = \Delta_0 e^{i \phi (r,t)}$), and the elastic energy of the phase fluctuations leads to interesting dynamics which manifests in the transport behaviour. In the \emph{strong pinning} regime, caused either by strong impurity potential or by dilute impurity concentration, the CDW phase is fully adjusted at each impurity site with the phase-phase correlation length being equal to the average impurity spacing. In the opposite limit of \emph{weak pinning} regime arising out of weak impurity potential or large impurity concentration, the CDW phase varies slowly, with its coherence length being larger than the average impurity spacing.

\begin{figure*}[ht]
\includegraphics[width=0.7\linewidth]{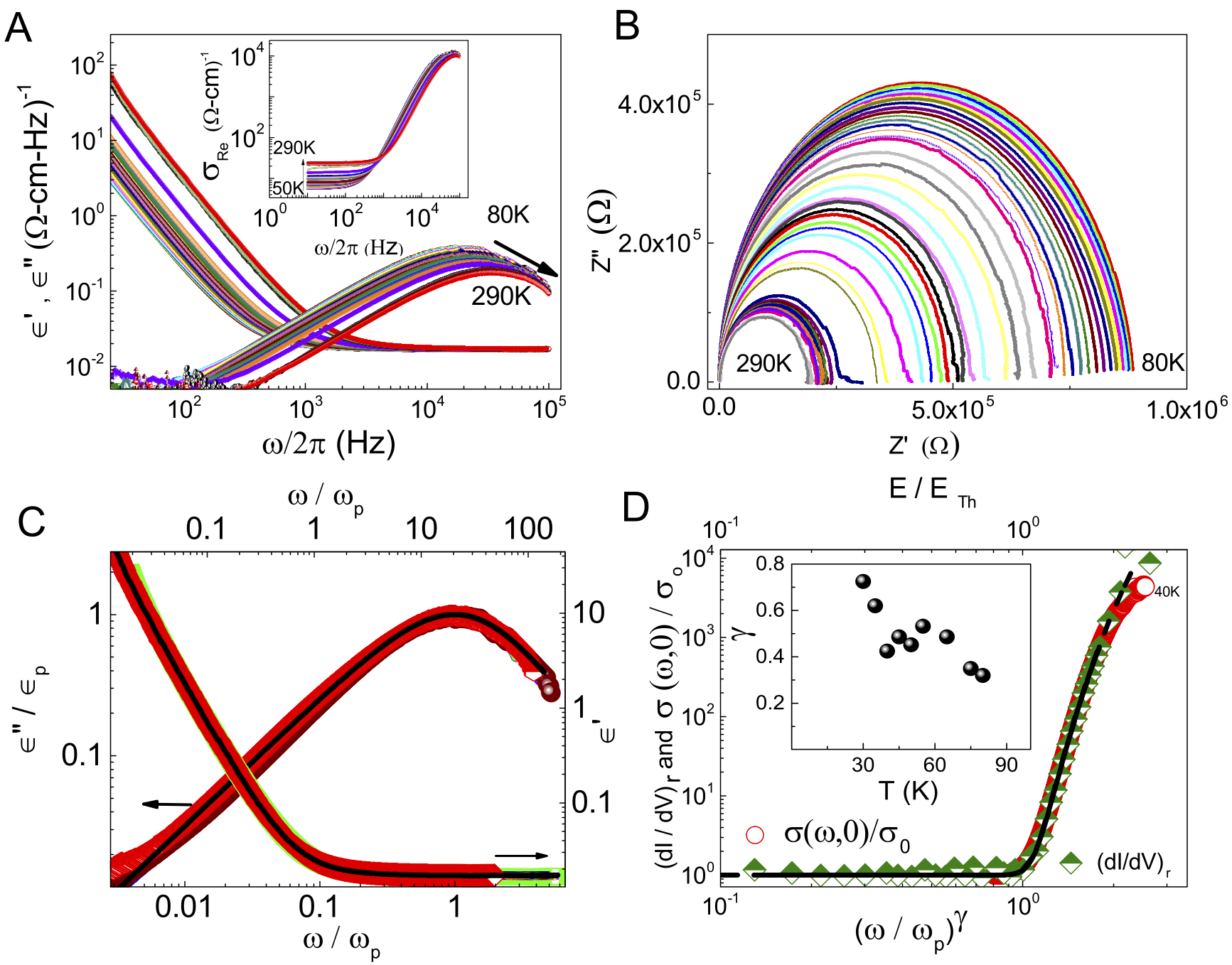}\\
\caption{Dielectric spectroscopy of YBIO for $T > 80$K: A) The frequency dependence of the real and imaginary part of the dielectric constant $\epsilon(\omega) = \epsilon'(\omega) + i \epsilon''(\omega) = [\sigma ({\omega}) - \sigma_0]/(i \omega)$, at different temperatures from 80K to 290 K. The plot
for $\epsilon''$ develops a peak at a $T$ dependent characteristic frequency $\omega_{\rm p} = 1/\tau(T)$.
The inset shows the real part of the conductivity at different temperatures. B) The corresponding Cole-Cole plots in terms of the impedance. C) The scaling of the dielectric response at different temperatures, on a universal generalized relaxation curve specified by Eq.~\eqref{acfit} -- represented by the solid black line. Here $\epsilon_{\rm p} = \epsilon''(\omega_{\rm p})$. D) The frequency dependence of the ac conductivity for YBIO scales with the voltage dependence of the dc conductivity at high temperature ($T \ge 80$ K) with suitable adjustment of the frequency scale defined by the exponent $\gamma$; The inset shows the temperature dependence of $\gamma$.}\label{fig3}
\end{figure*}

Bistable behavior and hysteresis in the low temperature dc I-V characteristics of CDW's, is a consequence of the elastic deformation of the CDW in the vicinity of pinning centers (impurities), for critical de-pinning field $E_{\rm T}$. This leads to the back-flow of the normal carriers screening the internal electric field \cite{Littlewood, Levy}, and is generally modeled by including the coupling between the quasi-particles and the CDW condensate in the FLR model \cite{Sneddon,Littlewood, Levy}.
The pinned and electrically driven CDW system first undergoes a transition into a plastically sliding state (at $E_{\rm T1}$), followed by a second hysteretic transition (at $E_{\rm T2}$) to a coherently moving CDW state \cite{Vinokur}. Alternately, the hysteretic behavior in dc transport of DW based systems can also
result from a phase slippage mechanism between weakly coupled CDW domains~\cite{Hall, Strogatz}.

Although both YBIO and YIO display similar dc characteristics such as hysteretic switching at low temperature, highly non-linear but smooth dc I-V characteristics at higher temperature, we find distinct difference between the two so far as the $T$ dependence of $E_{\rm T}$ is concerned. We find that for devices based on YIO, the threshold field increases sharply with decreasing temperature as T approaches zero (away from the CDW transition temperature), following an exponential dependence $E_{\rm T} \propto \exp(-T/T_0)$ (see inset of Fig.~\ref{fig2}B). On the other hand, $E_{\rm T}$ for devices based on YBIO is almost independent of temperature (inset of Fig.~\ref{fig2}A). The value of $E_{\rm T}$ is invariably higher in YIO based devices as compared to YBIO based ones of similar dimension. Such temperature dependence of $E_{\rm T}$ has also been observed earlier in standard charge and spin density wave systems~\cite{Flemming, 
SDW_Et} and is associated with thermally induced phase fluctuation and consequent renormalization of the condensate-impurity interaction~\cite{Maki}.

Another interesting feature of the dc conductivity is the cooling rate dependence of the CDW depinning field and the associated non-linear conduction and switching behaviour (see Fig.~\ref{fig2}C). This cooling rate dependence of $E_{\rm T}$ indicates that the pinning of CDW at low temperature is unlikely to arise form the boundary or surface of the sample or contact effects. The threshold $E_{\rm T}$ decreases with increasing cooling rate with eventual disappearance of the switching behaviour altogether leaving only smooth sliding transition (Fig.~\ref{fig2}C). Similar cooling rate dependence of CDW state which is otherwise absent in bulk crystals has recently been reported in low dimensional crystal of 1T-TaS$_2$~\cite{Yoshida}. The cooling rate affects the impurity layout as well as the domain growth. Rapid cooling might lead to shrinkage of domains thus increasing quasiparticle concentration at the domain boundaries, which in turn could influence the nature of depinning transition. Regarding the temperature dependance of the dc transport, we find that there is no change in the ohmic resistance below the threshold field with temperature in the low temperature regime. This suggests that at low temperature, the CDW is primarily pinned by the impurities and the interaction of CDW with quasi-particles is negligible.

We also studied the dc I-V characteristics of similar shaped YIO crystals of larger diameter $\sim 2\mu$m and found that the ohmic resistance below the threshold is still temperature independent up to $100$ K. However, the threshold electric field for switching decreases with increasing crystal radius (surface to volume ratio) similar to that reported in NbSe$_3$ earlier \cite{Carten}, and  the switching is always preceded by a non-ohmic regime unlike smaller crystals. If the lateral dimension of the crystal is larger than the transverse phase coherence length of the DW, then the DW can slide around local pinning centers leading to smooth de-pinning without switching. The fact that we still observe switching even for 2 $\mu m$ diameter system suggests that the CDW coherence length is significantly high ($\ge 2 ~\mu$m) in these systems.  Note that such large coherence length scales have also been reported earlier for other CDW systems such as TaS$_3$ and NbSe$_3$ \cite{Gruner_rmp,Gruner2}.

\subsection{Low frequency dielectric response}

Having described the DC transport in CDW's in YIO and YIBO, we now focus on their low frequency ($1-10^5$ Hz) ac response, up to room temperature. The frequency dependence of the real and the imaginary parts of complex conductivity for all the devices at different  temperatures was measured, using the standard lock-in technique.
In Fig.~\ref{fig3}A, we show the real and imaginary part of the low frequency dielectric response (related to the ac conductivity via the relation $i \omega \epsilon(\omega) = \sigma(\omega) - \sigma_0$, where $\sigma_0$ denotes the dc conductivity) for several temperatures (from 80 K to 290K). The corresponding Cole-Cole plot shows semicircular arc shape (Fig.~\ref{fig3}B) which suggests a Debye like dielectric relaxation mechanism. We find that in general the dielectric response at different temperatures, follows the generalized Debye relaxation formula (see Fig.~\ref{fig3}C) of Havriliak and Negami \cite{Havriliak,Degi}, which includes a skewed distribution of the dielectric relaxation rates and is given by
\begin{equation} \label{acfit}
\frac{\epsilon(\omega) - \epsilon_{\infty}}{\epsilon_0 - \epsilon_{\infty}} = \frac{1}{\left[1 + (i \omega \tau(T))^{1-\alpha}\right]^\beta}~.
\end{equation}
Here $\alpha$ denotes the temperature dependent width, and $\beta$ the skewness of the distribution of dielectric relaxation rates, and $\epsilon_{\infty} = \epsilon(\omega \to \infty)$.

Another well known experimental feature of the pinned CDW transport is the scaling between the dc and the ac conductivity \cite{Gruner2,Tucker2} at high temperatures for zero dc bias: ${\rm Re} \sigma_{\rm ac}(\omega)  = \sigma_{\rm dc}(V=\gamma \omega)$ where $\gamma$ is a frequency-voltage scaling parameter,
as shown in Fig.~\ref{fig3}D. While one plausible explanation for this scaling is generally given in terms of photon assisted quantum mechanical tunneling model \cite{Tucker2}, this ac/dc conductivity scaling behaviour has also been demonstrated numerically for a classical model based on phase pinning \cite{Sneddon1}.
We find the dc/ac scaling relation to be applicable (for $T \ge 80$ K), over a wide range of frequency (with some deviation at high frequency/voltage), with the scaling parameter $\gamma$ being (almost) inversely proportional to the temperature as shown in the inset of Fig.~\ref{fig3}D.
Similar behaviour for the dielectric relaxation at high temperatures (where there is no switching and no hysteresis) in CDW's (and SDW's) has also been seen earlier in K$_{0.3}$MoO$_3$ \cite{Cava,Degi}, (TaSe$_4$)$_2$I \cite{Cava1}, polypyrole nanotubes \cite{Sanyal} and Sr$_{14}$Cu$_{24}$O$_{41}$ ladder compounds \cite{Blumberg}.

\begin{figure}
\includegraphics[width=0.99\linewidth]{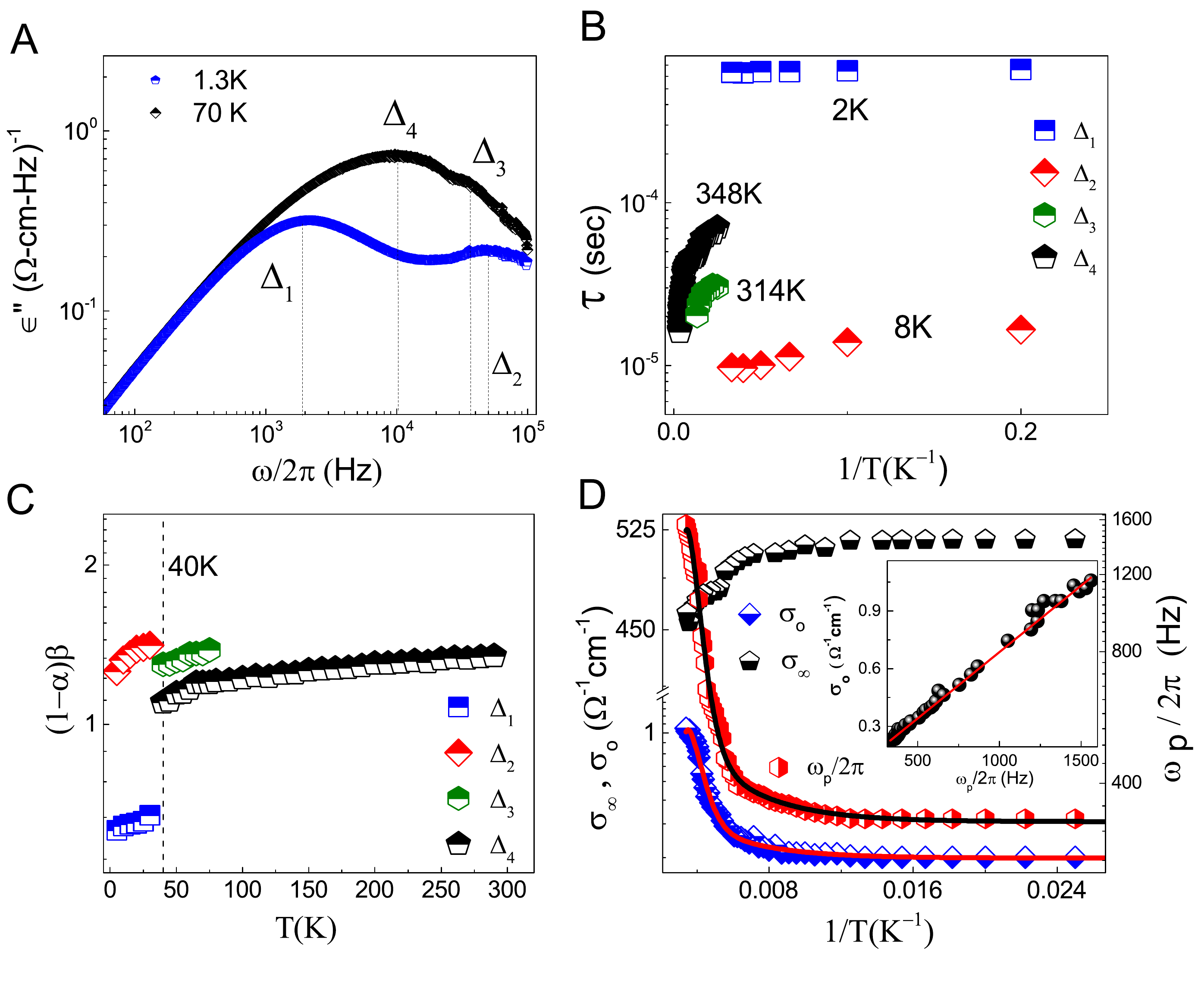}\\
\caption{A), B), C) Dielectric spectroscopy of YBIO at low $T$. A) Appearance of additional relaxation peaks in YBIO in the $\epsilon''$ versus $\omega$ plot as the temperature is lowered, is shown. B) Overall, there are four different relaxation rates, characterized by four different $\tau_i$, which follow Arrhenius $T$ dependence with the transition temperatures given by $\{T_1,T_2,T_3,T_4\} = \{ 2, 8, 314, 348\}$K. C) The corresponding temperature dependence of the product of the width ($1-\alpha_i$) and the skewness $(\beta_i)$ of the distribution of the four relaxation rates. D) The $T$ dependence of the dc ($\sigma_0$) and the high frequency
($\sigma_\infty$) conductivity, along with the prominent relaxation frequency ($\omega_{\rm p}$) as a function of the inverse of temperature in the whole temperature range up to room temperature.
The solid black (red) line shows the fit for $\omega_{\rm p}$ ($\sigma_0$) obtained by combining the Arrhenious behavior of the four scattering rates $\tau_i$ with the corresponding transition temperatures $T_i$. The dc conductivity follows the activated behaviour of the relaxation frequency over a wide range of temperature and frequency, as evident from the linear relationship between $\sigma_0$  and $\omega_{\rm p}$ shown in the inset.}\label{fig4}
\end{figure}

As $T$ is lowered below $80$ K, $\epsilon''(\omega)$ develops an additional peak at higher frequency, marked as $\Delta_3$ in Fig.~\ref{fig4}A. Now this additional peak can also be
characterized by Eq.~\eqref{acfit} though with a different $T$ dependence of 1) the characteristic scattering time, 2) the relaxation time distribution parameters: $\alpha$, $\beta$ and, 3) static dielectric constant ($\epsilon_0$). Further reduction of $T$ below $40$ K leads to the identification of two additional relaxation process -- four in all (labeled as $\Delta_i$ with $i = 1,2,3,4$) -- within the experimental frequency range, $10-10^5$ Hz (see Fig.~\ref{fig4}A). The temperature dependence of the $\tau_i$ and $(1-\alpha_i) \beta_i$ for all the four different relaxation processes is shown in Fig.~\ref{fig4}B and C, respectively. The scattering time of all the four different relaxation processes follow an Arrhenius temperature dependence ($\tau_i \propto \exp[-T_i/T]$) with a transition temperature of $\{T_1,T_2,T_3,T_4\} = \{ 2, 8, 314, 348\}$ K. The existence of the two low temperature scattering/relaxation processes (with $T_2 = 2$K, and $T_4 = 8$K), which become evident only for $T < 40$K are likely to be related to the two bistable phases at low temperature, which results in hysteresis in the dc transport characteristics (see Fig.~\ref{fig2}).
Similar behaviour has been observed earlier for CDW's in TaS$_3$ while exploring the full frequency and $T$ dependence of the dielectric response, and has been attributed to multi-phase glasslike phenomenology, strikingly similar to 3-fluoroaniline  \cite{Levy1}. For complete details of the temperature dependence of various parameters of Eq.~\eqref{acfit} for each of the four relaxation processes, see Ref.~\cite{SM}.

Figure \ref{fig4}D shows the temperature dependence of $\sigma_0$, $\sigma_{\infty}$ and the most prominent dielectric relaxation frequency $\omega_{\rm p}$. Note that $\omega_{\rm p} = \tau(T)^{-1}$  is also the peak frequency of the $\epsilon''(\omega)$ curve, and it serves as the frequency analogue of the threshold field $E_{\rm T}$. The prominent relaxation frequency $\omega_{\rm p}$ observed over the entire temperature regime is fitted well by combining the Arrhenious behaviour of the four scattering rates $\tau_i$ with the corresponding transition temperatures $T_i$.
Further it is evident from the inset of Fig.~\ref{fig4}D, that $\omega_{\rm p}(T) \propto \sigma_{\rm 0}(T)$. This is a consequence of the longitudinal damped collective charge oscillations in a CDW (polarization built up at the vicinity of the pinning sites), being screened by thermally excited `normal' quasi-particles \cite{Blumberg,Sanyal,Tucker1}.
In fact, we find that $\omega_{\rm p}/\sigma_0 \approx 0.98 \times 10^4$ Hz $\Omega$ cm, and using the relation for the low frequency longitudinal response which yields $\omega_{\rm p}/\sigma_0 \approx 4 \pi \epsilon_{\rm vacuum}/ (\epsilon_0 - \epsilon_{\infty})$
\cite{Blumberg}, we obtain the relative dielectric constant of YBIO to be $\epsilon_0 \approx 1.4\times10^{10}$ ($\epsilon_\infty \ll \epsilon_0$) which is consistent with the numbers obtained from the ac data fit using Eq.~\ref{acfit} (see tables in the supplementary material \cite{SM}). 
The temperature dependence of $\sigma_\infty$ calculated from the frequency dependent conductivity (for YBIO) in the high frequency limit (see Fig.~\ref{fig4}B) is weakly metallic as expected.

\section{Summary}
We have presented strong experimental evidence that the ground states of YIO and YBIO are charge density waves. Unambiguous transport signatures of a conventional density wave have been observed: hysteretic switching at low temperature, orders of magnitude enhancement of dc conductivity in switching and non-switching I-V characteristics with electric field, scaling of field and frequency dependent conductivity, strong low frequency dielectric response and similar temperature dependence of $\sigma_0$ and $\omega_{\rm p}$. A detailed analysis of the low frequency ac response, reveals the presence of four different relaxation processes - all with an Arrhenius dependence of the relaxation rate, with different transition temperatures. Note that while the experimental data, including the high values of $E_T$ and transition temperature, point towards the existence of CDW phase, the existence of SDW phase at very low temperature cannot be completely ruled out.  In addition, due to the presence of strong spin orbit interaction in pyrochlore irridates,  the density wave may actually be a mixture of charge- and spin-density waves. 
Nevertheless, it seems that the classical FLR model reasonably explains the low as well as higher temperature transport behaviour, with impurity pinning influencing the CDW dynamics at low temperature and CDW quasiparticle interaction dominating at high temperature.

Regarding the origin of the CDW ground state in low dimensional Pyrochlore Iridates, it is not clear
whether the underlying mechanism is purely electronic and related to improved Fermi surface nesting or whether orbital effects play any role. Pyrochlore Iridates are also expected to host the Weyl semi-metallic phase~\cite{Wan, William} and the CDW ordered state is a natural consequence of spontaneously broken chiral symmetry in three dimensional WSM, and may carry signature of axion dynamics \cite{Axion-CDW}. Indeed, a magnetic-field-induced charge density wave at
the pinned wave vector connecting Weyl nodes with opposite chiralities in Pyrochlore Iridates has already been predicted ~\cite{Kaiyuyang}. Since low-dimensional systems lack structural inversion symmetry because of surfaces, chiral symmetry could be broken even without magnetic field. The dynamical axion field may manifest itself as phase fluctuations of the CDW order parameter which couple to the electromagnetic field. Therefore, investigating the properties of the density wave and particularly its interplay with electromagnetic field would be an effective way of establishing the WSM phase or the Axion Insulator (AI) phase in Pyrochlore Iridates. Further experiments are needed to ascertain the nature of coupling of the density wave to the magnetic field. Nonetheless, the experimental observation of the density wave ground state suggests that realization of WSM/AI phase in mesoscopic Pyrochlore Iridate is a distinct possibility.

\end{document}